\documentclass[twocolumn,showpacs,preprintnumbers,amsmath,amssymb,aps,superscriptaddress]{revtex4}
\usepackage[pdftex]{graphicx}
\usepackage{qsymbols}
\usepackage{graphicx}

\begin{document}

\title{Effects of finite probe size on self-affine roughness measurements}
\author{F.~Lechenault}%
\email{frederic.lechenault@cea.fr}
\affiliation{CEA, IRAMIS, SPCSI, Grp. Complex Systems \& Fracture, F-91191 Gif Sur Yvette, France}%
\affiliation{Laboratoire des Collo\"ides, Verres et Nanomat\'eriaux, Universit\'e Montpellier 2, CNRS, France}%
\author{G.~Pallares}%
\affiliation{Laboratoire des Collo\"ides, Verres et Nanomat\'eriaux, Universit\'e Montpellier 2, CNRS, France}%
\affiliation{CEA, IRAMIS, SPCSI, Grp. Complex Systems \& Fracture, F-91191 Gif Sur Yvette, France}%
\author{M.~George}%
\affiliation{Laboratoire des Collo\"ides, Verres et Nanomat\'eriaux, Universit\'e Montpellier 2, CNRS, France}%
\author{C.~Rountree}%
\affiliation{CEA, IRAMIS, SPCSI, Grp. Complex Systems \& Fracture, F-91191 Gif Sur Yvette, France}%
\author{E.~Bouchaud}%
\affiliation{CEA, IRAMIS, SPEC, Grp. Instability \& Turbulence, F-91191 Gif Sur Yvette, France}%
\author{M.~Ciccotti}%
\email{ciccotti@lcvn.univ-montp2.fr}
\affiliation{Laboratoire des Collo\"ides, Verres et Nanomat\'eriaux, Universit\'e Montpellier 2, CNRS, France}%

\date{\today}

\begin{abstract}
The roughness of fracture surfaces has been shown to
exhibit self-affine scale invariance for a wide variety of materials
and loading conditions. The range of scales over which this regime
extends remains a matter of debate, together with the universality of the associated exponents.
The topography of these
surfaces is however often investigated with a contact probe
that is larger than the micro-structure.
In this case, we show that the correlation function of the roughness and the corresponding Hurst exponent $\zeta$
can only be measured down to a
length scale $\Delta x_c$ which depends on the probe size $R$, on $\zeta$ and on the surface topothesy $l$, and exhibit spurious behavior at smaller scales. 
First, we derive the dependence of
$\Delta x_c$ on these parameters from a simple scaling argument.
Then we study this dependence numerically and verify our
theoretical prediction. Finally, we establish the relevance of this analysis from AFM measurements on an experimental glass fracture surface and provide a metrological procedure for roughness measurements. 
\end{abstract}


\maketitle

Fractography has long been a fruitful method to decipher the
microscopic mechanisms of fracture in heterogeneous materials (see
e.g.,\cite{Bouchaud1997,Bouchaud2003} for reviews). Since
an early paper by Mandelbrot et al.
\cite{Mandelbrot1984}, a growing interest has developed in the
understanding of the fractality of crack surfaces. One of
the most prominent features of these surfaces
is the self-affine structure of their roughness in various materials, ranging from ductile metallic
alloys to brittle mortars or glasses
\cite{Maloy1992,Daguier1996,rountree2002,Mourot2006,Bonamy2006,Santucci2007,Dalmas2008,Morel2008}.
A proper understanding and experimental characterization of this feature has both
applied and fundamental stakes: on the one hand, it has recently
led to a method to extract the direction of crack propagation from
the analysis of roughness, which is of high practical importance
\cite{Ponson2006}. On the other hand, the possible universality underlying this critical-like behavior remains to be assessed
\cite{Bouchaud1990}, together with the physical origin of the
intrinsic cutoffs in the self-affine regimes
\cite{Bonamy2006,Ponson2006}. These scaling
properties have motivated the development of numerous
models \cite{Hansen1991,Bouchaud1993,Schmittbuhl1995,Daguier1997,Ramanathan1997,Hansen2003,Nukala2006}.

\indent Recent advances in scanning probe microscopy (SPM),
especially the development of atomic force microscopy (AFM), have
provided large amounts of experimental data drawing a rather
complicated picture of the situation: the reported 
exponents depend on the direction of analysis, the material and the resolution. Furthermore, the characteristic length
scale of the self-affine regimes, or {\it topothesy}, is rarely
reported or unphysical.
This discrepancy in the reported exponents
can originate from the existence of several universality classes and actual differences in the underlying physical
mecanisms. However, it might also partly result from experimental
biases. The list of possible biases is long and instrument
dependent, i.e.~environment noise, feedback loops,
biases in the estimators \cite{Schmittbuhl1995}, etc. Here, we
focus on the subtle bias induced by the non-linear smoothing effect of the
finite-size contact probe used in topography measurements of self-affine surfaces
\cite{Mazeran2005}.

In order to quantitatively estimate this bias,
we investigate the consequences of using model probes of various sizes and shapes on ideal self-affine surfaces. As SPM
topography measurements usually proceed line by line, we will
study the one-dimensional problem of probing a self-affine profile
$h\left(x\right)$ with a rigid tip defined by
$f^{\left[\gamma\right]}_R\left(x\right)\equiv\frac{\left|x\right|^{\gamma}}{2R^{\gamma-1}}$, where $R$ is the characteristic size of the apex and $\gamma$ a shape parameter ($\gamma \geq 2$). This choice of shape is motivated by the possible flattening of real tips during scans. The roughness features of the profile $h$ are
extracted from the height-height correlation function defined as
\begin{equation}
\Delta h\left(\Delta x\right)\equiv \sqrt{\langle\left(h\left(x+\Delta x\right)-h\left(x\right)\right)^2\rangle_x}
\end{equation}
The profile $h\left(x\right)$ is self-affine if it is invariant under the transformation
\begin{equation}
x \rightarrow\lambda x \qquad h \rightarrow \lambda^{\zeta} h
\end{equation}
where $\zeta$ is the roughness or Hurst exponent.
It follows that there is a physical length scale $l$, called the {\it topothesy}, for which the local slope of the profile becomes one. It verifies
\begin{equation}
\frac{\Delta h}{l}=\left(\frac{\Delta x}{l}\right)^{\zeta}\quad\mathrm{i.e}\quad\Delta h=\Delta x^{\zeta}l^{1-\zeta}
\label{eqn:topo}
\end{equation}
Physically, this parameter provides information about the
amplitude of the roughness. In particular, it follows from
Eq.~(\ref{eqn:topo}) that vertically dilating the profile affects the
topothesy as
\begin{equation}
h \rightarrow\mu h \qquad l \rightarrow\mu^{\frac{1}{1-\zeta}} l
\label{eqn:zoom}
\end{equation}

\begin{figure}[h]
\vspace{-0.4cm}
\includegraphics[width=0.5\textwidth]{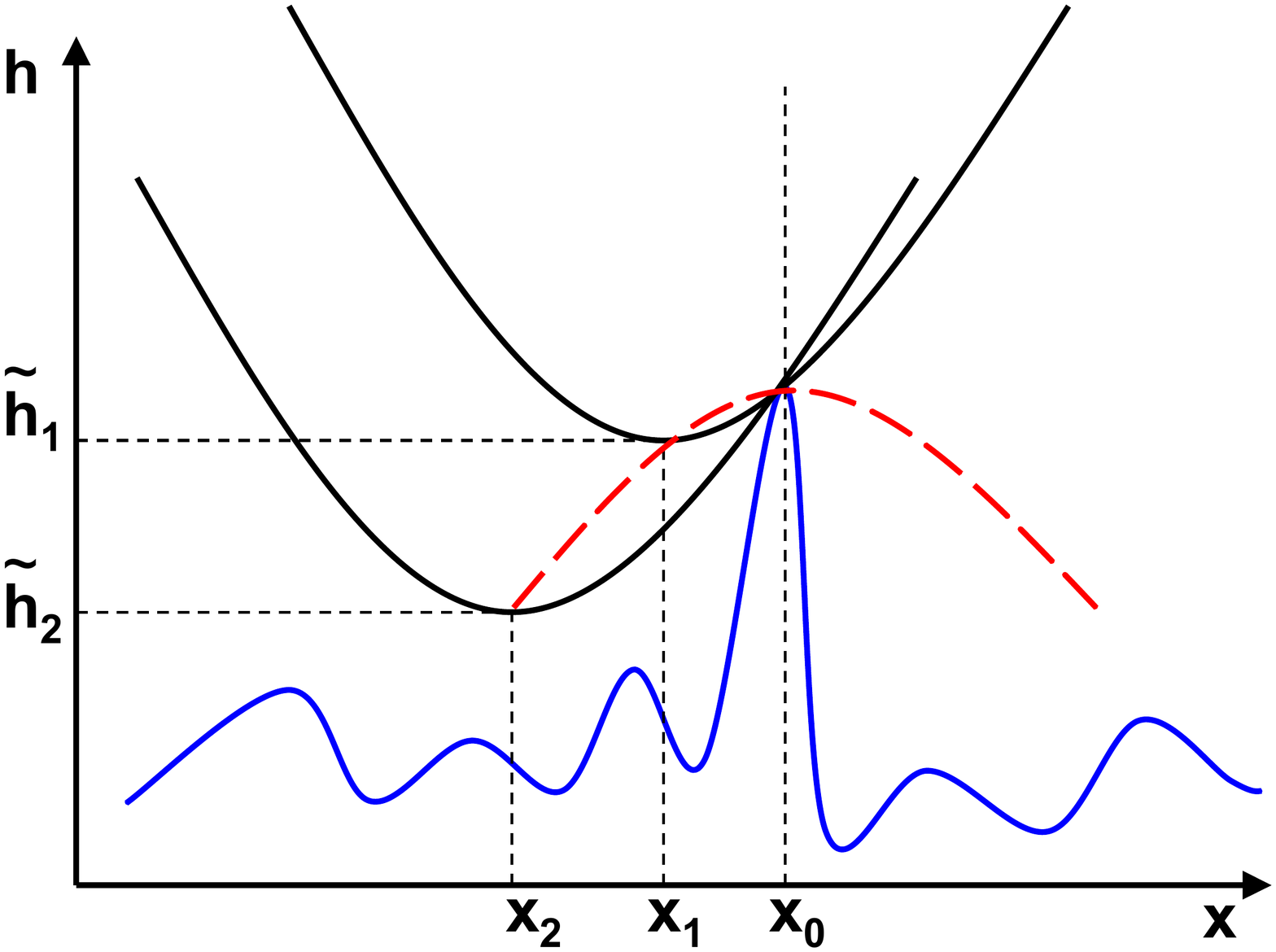}
\vspace{-0.7cm}
\caption{Schematic of the small scale behavior of the smoothed profile. For two closeby points, the probe (black) touches a unique distant peak on the profile (blue). We have drawn the corresponding portion of the measured profile $\tilde{h}$ (dotted red)}.
\label{fig:schema}
\end{figure}

Measuring such a profile with a finite-size SPM tip amounts to finding the height, $\tilde{h}\left(x\right)$ of the apex as the tip first touches the surface when approached from above, for each value of $x$, as depicted in Fig.~\ref{fig:schema}. This occurs at some position $x_0$ such that
\begin{equation}
\tilde{h}(x) \equiv  {\cal S}_R\left[ h \right]\left( x \right) \equiv \textrm{max}_{x'} \left(h\left(x'\right) - g_R\left(x-x'\right)\right)
\label{eqn:smoothing}
\end{equation}
where $g_R\left(x\right)$ describes the shape of the tip, $R$ its size, and ${\cal S}_R$ is a smoothing operator acting on profiles. 
${\cal S}_R$ is non-linear $\left({\cal S}_R\left[ \mu h \right] \neq \mu {\cal S}_{R}\left[ h \right]\right)$ and implies partial loss of information. Interestingly, for tip shapes $g_R=f^{\left[\gamma\right]}_R$, we can still derive a scaling relation for ${\cal S}_R\left[ h \right]$. Writing Eq.~(\ref{eqn:smoothing}) for a profile scaled according to Eq.~(\ref{eqn:zoom}) yields
\begin{eqnarray}
{\cal S}_R\left[ \mu h \right] &=& \textrm{max}_{x'} \left(\mu h(x') - \frac{\left|x-x'\right|^{\gamma}}{2R^{\gamma-1}}\right)\nonumber\\
{\cal S}_R\left[ \mu h \right] &=& \mu \left\{ \textrm{max}_{x'} \left(h(x') - \frac{\left|x-x'\right|^{\gamma}}{2\left(\mu^{\frac{1}{\gamma-1}} R\right)^{\gamma-1}}\right) \right\}\nonumber\\
{\cal S}_R\left[ \mu h \right] &=& \mu {\cal S}_{R'}\left[ h \right]\quad\textrm{with}\quad R'=\mu^{\frac{1}{\gamma-1}} R
\label{eqn:equiv}
\end{eqnarray}
This relationship states that the bias induced by scanning a profile with a given tip-radius is equivalent to that obtained by scanning the vertically expanded profile with a thinner tip.
Now we turn to the investigation of the influence of the local probe size on the height-height correlation function. Qualitatively, applying ${\cal S}$ dampens the roughness details at length scales smaller than a characteristic length $\Delta x_c$. Let us consider two nearby points $x_1$ and $x_2$ close enough to a point $x_0$ such that $\tilde{h}\left(x_{1,2}\right)=h\left(x_0\right)-f^{\left[\gamma\right]}_R\left(x_{1,2}-x_0\right)$, as depicted in Fig.~\ref{fig:schema}. Setting $x_2=x_1+\Delta x$, for values of $\Delta x$ much smaller than $\Delta x_c$, we get $\tilde{h}\left(x_2\right)-\tilde{h}\left(x_1\right)=f^{\left[\gamma\right]}_R\left(x_1-x_0+\Delta x\right)-f^{\left[\gamma\right]}_R\left(x_1-x_0\right)\sim \frac{df^{\left[\gamma\right]}_R}{dx}\left(x_1-x_0\right)\Delta x$. The root mean square of this quantity, $\Delta\tilde{h}$, scales like $\Delta x$, with a prefactor that we conjecture to be proportional to $\frac{df^{\left[\gamma\right]}_R}{dx}\left(\Delta x_c\right)$.
On the other hand, the very large scale structure of the profile is not affected, and we expect $\Delta \tilde{h} \sim \Delta x^{\zeta}$ for $\Delta x \gg \Delta x_c$. We can now evaluate how $\Delta x_c$ scales with $R$, ${\zeta}$ and $l$ by assuming that at $\Delta x_c$, the height of the tip equates the height-height correlation:
\begin{equation}
f^{\left[\gamma\right]}_R\left(\Delta x_c\right) \sim \Delta h\left(\Delta x_c\right)
\end{equation}
Using Eq.~(\ref{eqn:topo}) and the expression for the chosen tip shape yields
\begin{equation}
\frac{\Delta x_c^{\gamma}}{2R^{\gamma-1}} \sim \Delta x_c^{\zeta} l^{1-{\zeta}}
\end{equation}
Solving for $\Delta x_c$ and dropping the numerical prefactor gives the approximate scaling relation
\begin{equation}
\Delta x_c \sim \left(R^{\gamma-1}l^{1-\zeta}\right)^{\frac{1}{\gamma-\zeta}}
\label{eqn:scaling}
\end{equation}

\begin{figure}[h]
\includegraphics[width=0.9\linewidth]{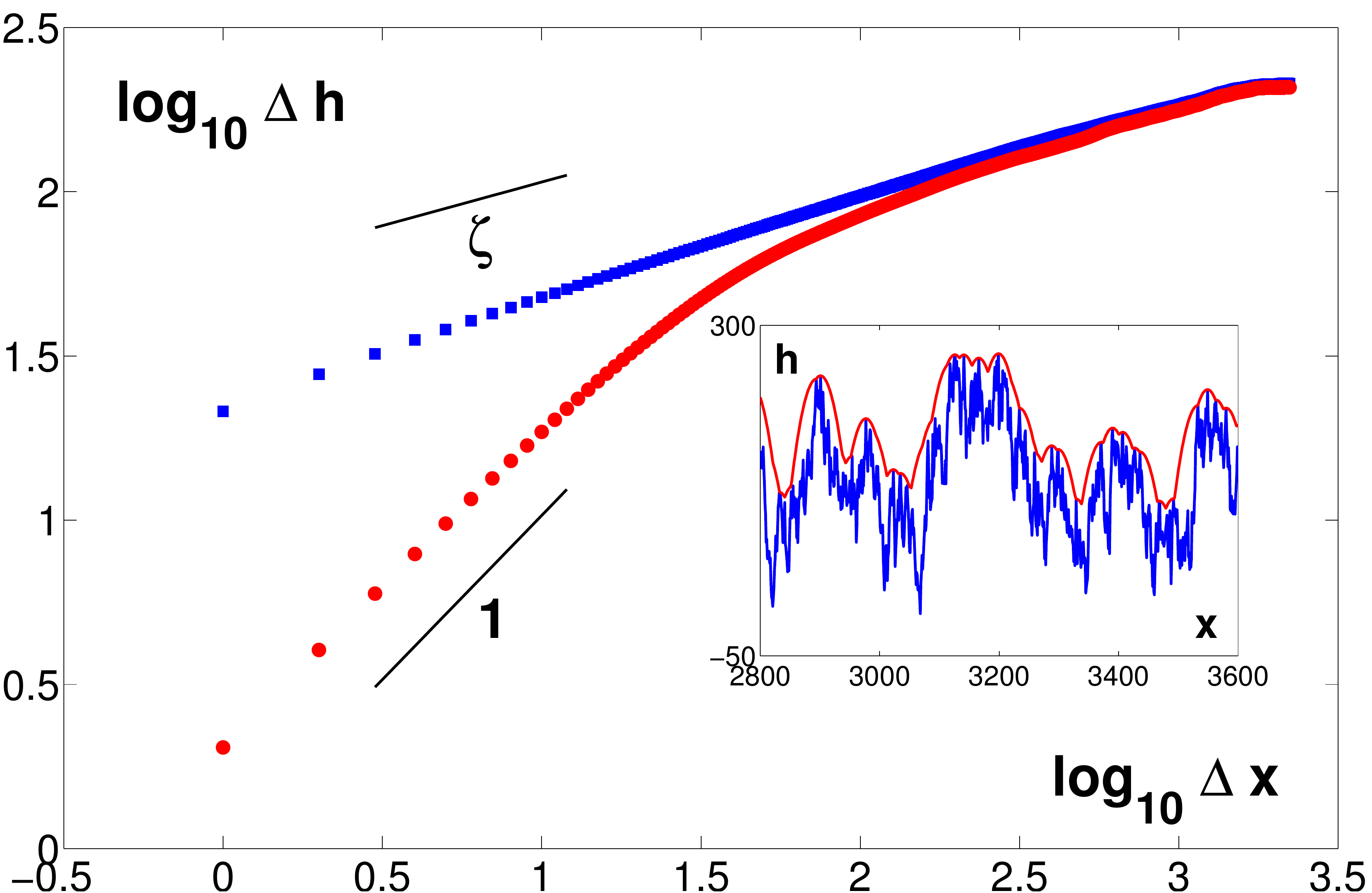}
\caption{{\bf Main panel:} Height-height correlation function evaluated for i) a self-affine profile with ${\zeta}=0.33$ and $l=100$ ($`(!)$), ii) the same profile smoothed with a tip of size $R=7.8$ ($`[!]$). {\bf Inset:} Portion of the corresponding original (blue) and smoothed (red) profiles.}
\label{fig:HH}
\end{figure}

In order to assess the validity of this relation, we have performed numerical simulations of self-affine profiles with exponents ${\zeta}$ ranging from 0.2 to 0.9 and scaled the topothesy $l$ in the range $10^{-3} l_0$ to $10^3 l_0$ according to Eq.~(\ref{eqn:zoom}). Due to its high representativeness, we focus on the parabolic case $\gamma=2$ with tip radius of curvature $R$ ranging from $10^{-5} l_0$ to $10^3 l_0$, where $l_0$ is the spatial sampling interval. In the following, we will drop $l_0$ in our notations, and consider all length scales in units $l_0$. The self-affine profiles have been obtained using the spectral method described in \cite{Voss1988}; they contain $20000$ data points each. For all values of these parameters, we compute the height-height correlation function for the original and the smoothed profiles. An example is shown in the main panel of Fig.~(\ref{fig:HH}). The self-affine exponent and topothesy of the simulated profiles are evaluated according to the following method. We define the function $\Lambda$ by
\begin{equation}
\Lambda\left(\Delta x, \eta \right) = \left(\frac{\Delta h}{\Delta x^{\eta}}\right)^{\frac{1}{1-\eta}}
\end{equation}
It follows from Eq.~(\ref{eqn:topo}) that $\Lambda\left(\Delta x, {\zeta} \right)=l$ for all $\Delta x$. We evaluate $\Lambda\left(\Delta X, \eta \right)$ for a uniform random variable $\Delta X$. ${\zeta}$ is then defined as the value for which the variance of $\Lambda$ reaches a minimum in $\eta$, and the average value of $\Lambda\left(\Delta X, {\zeta} \right)$ provides an estimator of $l$ based on its fundamental invariance property $\Delta h\left(l\right)=l$. 

\begin{figure}[h]
\vspace{-0.2cm}
\includegraphics[width=0.45\textwidth]{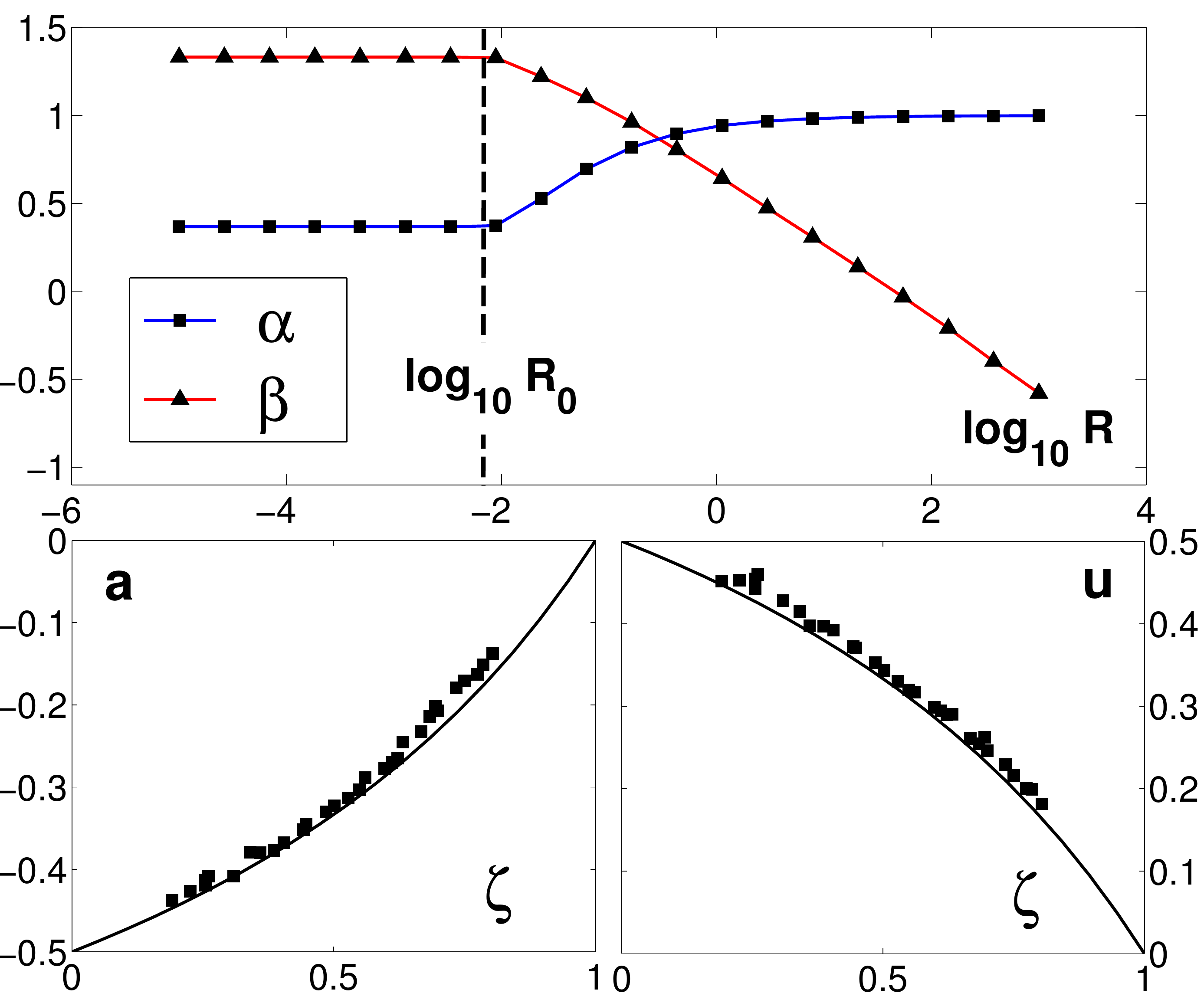}
\caption{{\bf Top:} Apparent parameters $\alpha$ and $\beta$ as a function of $\log_{10} R$ for ${\zeta}=0.33$ and $l=100$. {\bf Bottom left:} Parameter $a$ as a function of exponent $\zeta$. The line represents $-\frac{1-{\zeta}}{2-{\zeta}}$. {\bf Bottom right:} Parameter $u$ as a function of exponent $\zeta$. The line represents $\frac{1-{\zeta}}{2-{\zeta}}$.}
\label{fig:AlphaBeta}
\end{figure}

We observe a clear crossover between the small and large scale behavior of $\Delta \tilde{h}$. At large scales,  $\Delta \tilde{h}$ asymptotically behaves like $\Delta h$. It has to be noted that it reaches this regime very slowly, and even though the exponent reaches $\zeta$ from above, $\Delta \tilde{h}$ remains underneath the original correlation function, resulting in underestimation of the surface roughness amplitude and overestimation of the exponent. At scales smaller than the crossover length scale $\Delta x_c$, $\Delta \tilde{h}$ bends down and reaches the expected linear scaling after a long transient. In order to extract the position of the cutoff $\Delta x_c$, we extract the local behavior of $\Delta \tilde{h}$ around $\Delta x = l_0$ by fitting a line
\begin{equation}
\log_{10}\Delta \tilde{h} = \alpha(R,{\zeta},l) \log_{10}\Delta x + \beta(R,{\zeta},l)
\end{equation}
through the first two points of $\Delta \tilde{h}$: $\alpha$ is essentially the log-derivative of $\tilde{h}$ at the resolution $l_0$ (i.e.~at the smallest accessible value of $\Delta x$). 
We have represented in Fig.~\ref{fig:AlphaBeta}(Top) the behavior of $\alpha$ and $\beta$ as a function of the tip radius $R$ for ${\zeta}=0.33$ and $l=100$. Two regimes are clearly observed. For values of $R$ smaller than a cutoff $R_0 ({\zeta},l)$, the height-height correlation function is not affected by the smoothing effect. From Eq.~(\ref{eqn:topo}), we get that in this regime, $\alpha (R,{\zeta},l) = {\zeta}$ and $\beta \left(R,{\zeta},l\right) = \left(1-{\zeta}\right) \log_{10} l$ when $R < R_0({\zeta},l)$. For values of $R$ much larger than $R_0$, $\alpha$ converges towards $1$, as expected, whereas $\beta$ essentially behaves like a power law in $R$.
In order to extract the scaling of the threshold $R_0$, we fit $\beta = a\left({\zeta},l\right)\log_{10} R +b\left({\zeta},l\right)$. The function $a\left({\zeta},l\right)$, appears to depend only marginally on $l$, which we neglect. The dependence of $a$ on $\zeta$ is depicted in Fig.~\ref{fig:AlphaBeta} (Bottom left). It can be accurately modeled by $a = -\frac{1-{\zeta}}{2-{\zeta}}$.
We now turn to the scaling of $b$, which is well described by $b = u\log_{10} l +v$. No significant dependence on ${\zeta}$ is found for $v$, and this number is of the order of $-0.1$, which is negligible and will be dropped henceforth. We find $u = \frac{1-{\zeta}}{2-{\zeta}}$, as shown in Fig.~\ref{fig:AlphaBeta} (Bottom right). Altogether, the behavior of $\beta$ conveniently writes
\begin{equation}
\beta = \left(\frac{1-{\zeta}}{2-{\zeta}}\right)\log_{10} \frac{l}{R} \quad \textrm{when} \quad R \gg R_0
\end{equation}
We can now extract $R_0$ by intersecting the small and large scale behaviors of $\beta$, which reads, in units $l_0$
\begin{equation}
\left(1-{\zeta}\right)\log_{10} l = \left(\frac{1-{\zeta}}{2-{\zeta}}\right)\log_{10} \frac{l}{R_0}
\end{equation}
This yields
\begin{equation}
\log_{10}R_0=\left({\zeta}-1\right)\log_{10} l
\end{equation}
It is interesting to notice that since ${\zeta}-1$ is smaller than $1$, the effect of the tip appears for smaller radii when the topothesy is larger. This scaling is a consequence of the equivalence expressed in Eq.~(\ref{eqn:equiv}). The existence of this cut-off is an effect of the resolution $l_0$ of our measurement. As a matter of fact, restoring the sampling length scale yields $\Delta x_c\left(R_0\right)=l_0$. This means that for radii smaller than $R_0$, the cut-off length $\Delta x_c$ becomes smaller than the resolution and hence disappears from the resolved portion of the correlation function.
In the asymptotic regime where $R \gg R_0$, the small scale exponent $\alpha$ goes to $1$, as can be seen in Fig.~\ref{fig:AlphaBeta}(Top). Thus the asymptotic form of the correlation function for $R\gg R_0$, or equivalently $\Delta x_c\gg \Delta x$, reads:
\begin{equation}
\log_{10}\Delta \tilde{h} = \log_{10}\Delta x +\frac{1-{\zeta}}{2-{\zeta}}\log_{10}\frac{l}{R}
\end{equation}
Solving $\Delta \tilde{h}\left(\Delta x_c\right)=\Delta h\left(\Delta x_c\right)$ yields Eq.~(\ref{eqn:scaling}) derived above for the expression of $\Delta x_c$.
This result confirms our asymptotic conjecture in the parabolic case since 
\begin{equation}
\log_{10}\frac{df^{\left[2\right]}_R}{dx}\left(\Delta x_c\right)=\log_{10}\frac{\Delta x_c}{R}=\beta
\end{equation}

\begin{figure}[h]
\includegraphics[width=0.45\textwidth]{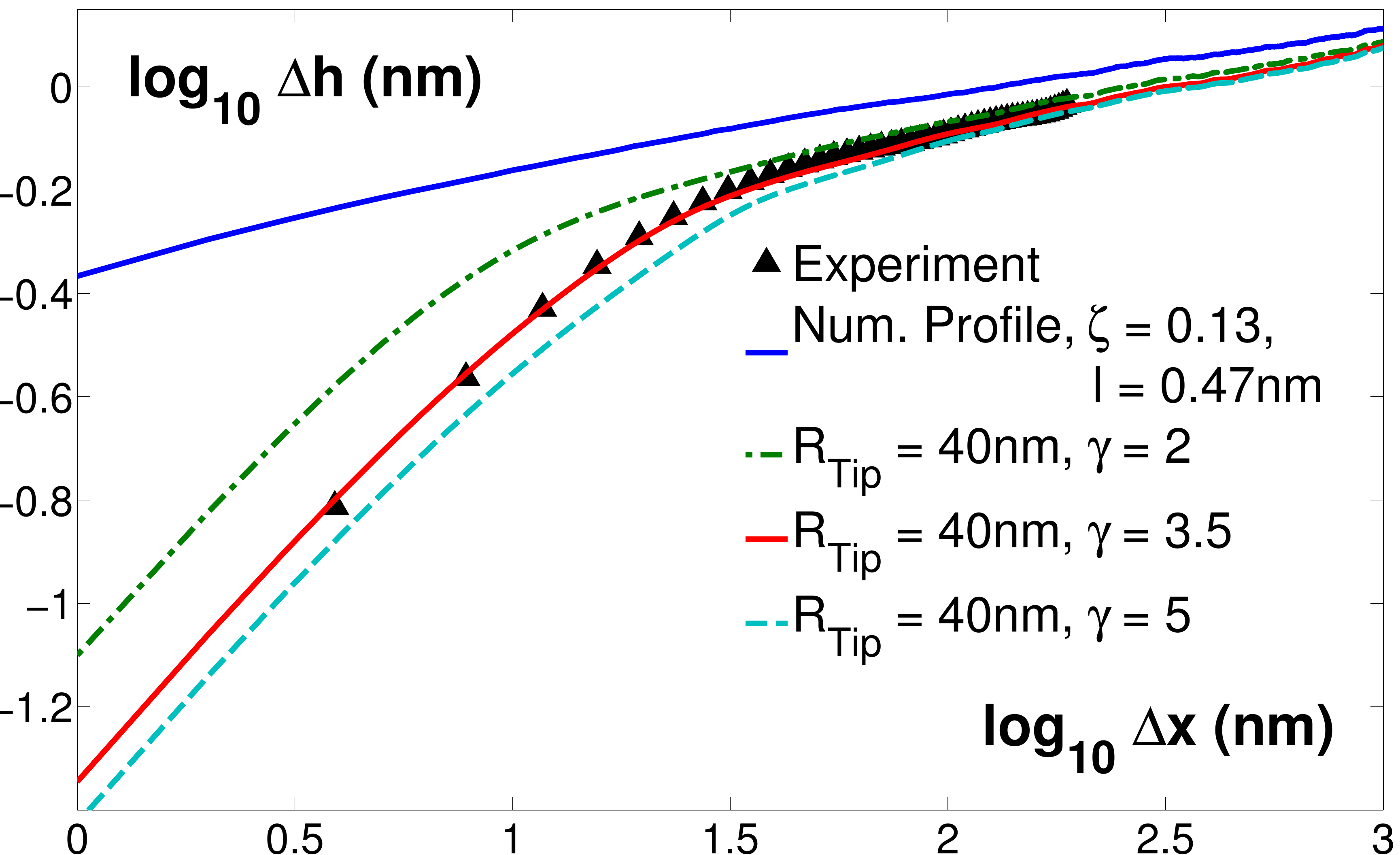}
\caption{Experimental height-height correlation functions obtained for a silica glass fracture surface scanned with an AFM with tip radius $R=40nm$. Original and smoothed height-height correlation functions obtained for a numerical profile with $\zeta=0.13$ and a $40nm$ tip for different values of $\gamma$.}
\label{fig:ExpVSModel}
\end{figure}

To check the relevance of this analysis, we have performed AFM (Veeco Nanoscope Dimension V) measurements on fused silica glass (Corning 7980) fracture surfaces. We break a DCDC sample in the stress-corrosion regime, in conditions detailed in \cite{Bonamy2006}. Then we scan the topography of the fracture surface with the AFM in tapping mode ($2\times 2\mu m^2, 512\times 512$ pixels). 
The height-height correlation function of the data parrallel to the crack front is shown in Fig.~\ref{fig:ExpVSModel}. Two scaling regimes are observed: a $0.8$ exponent at small scales, and a $0.2$ exponent at large scales. 
Once this is done, we acquire an image of the tip of the AFM with a silicon tip characterizing grating (TGT01, NT-MDT, Russia). 
By fitting $f^{\left[\gamma\right]}_R$ on this measurement, we evaluate $R_{Tip}=40\pm5nm$ and a shape parameter $\gamma=4\pm0.5$, which is typical for AFM tips on hard substrates. 
Then, through trial and error, we select a numerical profile with $\zeta=0.13$, which is close to the classical logarithmic behavior \cite{Ramanathan1997}. We smooth it using the measured value of $R_{Tip}$, and different values of $\gamma$. The resulting correlation functions for this value of $\zeta$, superimposed to the experimental ones in Fig.~\ref{fig:ExpVSModel}, reproduce quantitatively our measurements for $\gamma=3.5$, which is compatible with the measured shape of the tip. 

Altogether, these results strongly suggest that the small scale self-affine regime with a large exponent and the associated cut-off we observe on silica glass fracture surfaces are due to the tip-smoothing effect. We conjecture that earlier measurements using contact probes on different materials might exhibit such bias. The numerical procedure we proposed provides a sharp metrological constraint on roughness characterization and allows for accuracy checks on earlier results. Further work is needed to understand the slow convergence of the correlation function at large scales and reconstruct the statistical properties of the original roughness from a smoothed profile. 
         
To conclude, we have shown that the use of a finite-size contact probe to scan a self-affine surface significantly affects its measured small scale roughness properties. In particular, we have derived theoretically and checked numerically the dependence of the length scale at which this effect starts on the surface parameters and the asymptotic small scale behavior of the height-height correlation function. Finally, we have performed experiments confirming the relevance of this analysis regarding AFM metrology and provided insight into the origin of previously reported large Hurst exponents at small scales. 

We thank D. Bonamy and A. Grimaldi for fruitful discussion. All co-authors of this work wish to acknowledge the financial support of ANR Grant ``Corcosil'' No. ANR-07-BLAN-0261-02.

\bibliographystyle{apsrev}

\end{document}